\documentclass{article}

\usepackage{graphicx,amssymb,amsmath,amsfonts,mathtext,color,wrapfig}
\usepackage{cite,enumerate,float}
\usepackage{color}
\usepackage{tikz}
\usetikzlibrary{arrows,snakes,backgrounds}

\usepackage[utf8]{inputenc}

\def\be{\begin{eqnarray}}
\def\ee{\end{eqnarray}}

\definecolor{red}{rgb}{1,0,0}
\definecolor{orange}{rgb}{1,0.5,0}
\definecolor{violet}{rgb}{0.7,0,1}

\hoffset= - 1.0in         % switch off for draft style
\begin{document}

% \hfill ??? \today
\hfill MIPT/TH-13/25

\hfill FIAN/TH-20/25

\hfill IITP/TH-14/25

\hfill ITEP/TH-16/25

\bigskip

\centerline{\Large{
\textbf{5D AGT conjecture for circular quivers}
}}

\bigskip

\centerline{A. Mironov$^{b,c,d,e}$, A. Morozov$^{a,c,d,e}$, Sh. Shakirov$^d$}

\bigskip

\begin{center}
$^a$ {\small {\it MIPT, Dolgoprudny, 141701, Russia}}\\
$^b$ {\small {\it Lebedev Physics Institute, Moscow 119991, Russia}}\\
$^c$ {\small {\it NRC ``Kurchatov Institute", 123182, Moscow, Russia}}\\
$^d$ {\small {\it Institute for Information Transmission Problems, Moscow 127994, Russia}}\\
$^e$ {\small {\it Institute for Theoretical and Experimental Physics, 117218, Moscow, Russia}}
\end{center}

\bigskip

\centerline{ABSTRACT}

\bigskip

{\footnotesize
The best way to represent generic conformal blocks
is provided by the free-field formalism,
where they acquire a form of multiple Dotsenko-Fateev-like integrals
of the screening operators. Degenerate conformal blocks can be described
by the same integrals with special choice of parameters. Integrals satisfy various recurrent relations,
which for the special choice of parameters reduce to closed equations. This setting is widely used in explaining the AGT relation,
because similar integral representations exist also for Nekrasov functions.
We extend this approach to the case of $q$-Virasoro conformal blocks
on elliptic surface — generic and degenerate. For the generic case, we check equivalence
with instanton partition function of a 5d circular quiver gauge theory. For the degenerate case, we check equivalence
with partition function of a defect in the same theory, also known as the Shiraishi function.
We find agreement in both cases. This opens a way to re-derive the sophisticated equation for the Shiraishi function as the equation for the corresponding integral, what seems straightforward, but remains technically involved and is left for the future.
}

\bigskip

\bigskip

\section{Introduction}

The celebrated AGT conjecture \cite{AGT} relates instanton partition functions in four-dimensional ${\cal N} = 2$ supersymmetric Yang-Mills gauge theory with conformal blocks of two-dimensional Liouville conformal field theory. Concrete realizations of this relation vary, because there exist several equivalent descriptions: instanton partition functions can be represented as LMNS multiple contour integrals \cite{LMNS1,LMNS2,LMNS3} or as Nekrasov sums over partitions (Young diagrams) \cite{Nekrasov}.
Likewise, two-dimensional conformal blocks can be given with the help of operator product expansion \cite{OPE,OPEhigher} or with the help of free-field formalism as Dotsenko-Fateev integrals (matrix models) \cite{DF,DFnew}. Those different descriptions are useful in different contexts and highlight complementary properties of the problem.

Among other descriptions, the formalism of Dotsenko-Fateev (DF) integrals has considerable advantages  \cite{DFnew}.
For example, its explicit choice of integration contours allows one to describe   conformal blocks at finite values of all parameters.
It also makes direct contact with well-developed theory of matrix models, including Dijkgraaf-Vafa phases \cite{DV} and exact Selberg-Kadell formulas for correlators \cite{Selb,Kadell}.
Last but not the least, it provides a clear understanding of the AGT conjecture as a Hubbard-Stratonovich duality \cite{HS,MMNek} of matrix models under two alternative expansions of their integrand.

Since its discovery, it was clear that AGT conjecture admits many generalizations: some straightforward, some more puzzling. A few important directions of generalization are:

\paragraph{$\bullet$ $q$-deformation.}
On the Yang-Mills side  this is a lift from four-dimensional to five-dimensional gauge theory \cite{5dAGT,5dAGT2}.
On the conformal field theory side, this corresponds to $q$-deformation, replacing basic rational functions in conformal field theory by their trigonometric counterparts. Sometimes there exists an even further \begin{wrapfigure}{r}{0.3\textwidth}
{\vspace{-6ex}
  \begin{center}
    \includegraphics[width=0.25\textwidth]{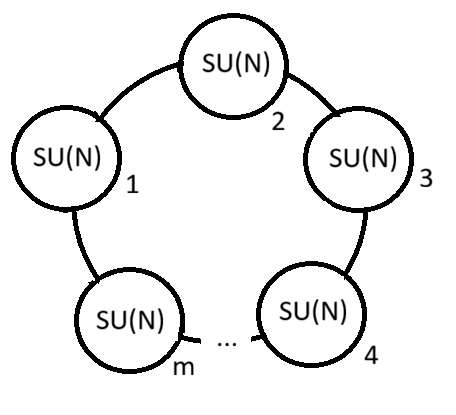}
  \end{center}
  \vspace{-6ex}}
\caption{Circular quiver of $m$ gauge groups of type $SU(N)$.}
  \vspace{-5ex}
\end{wrapfigure}six-dimensional generalization, which corresponds to replacing trigonometric functions by elliptic functions.

\paragraph{$\bullet$ Higher genus.}
This implies replacing the sphere, where conformal field theory is defined, by a torus
or higher genus surface. On the gauge theory side, this corresponds to more complicated quivers of gauge groups. In particular, conformal blocks on a torus correspond to a circular quiver as shown on Figure 1.
Note that in this way elliptic functions naturally  appear in the problem, however this elliptization is different from the one associated to six-dimensional generalization, and interplay of the two remains a topic of active research \cite{DELL,MMZ,MMdell}.

\paragraph{$\bullet$ Higher rank.}This deals with gauge groups of higher rank such as $SU(N)$ as opposed to the rank one group $SU(2)$. On the conformal field theory side, this corresponds to promoting Liouville conformal field theory to Toda theory where the Virasoro symmetry is enhanced to a ${\cal W}_N$ algebra \cite{WnAGT,MMWn}.

\paragraph{$\bullet$ Including defects.}This has to do with instanton partition functions in the presence of surface defects in the gauge theory. On the conformal field theory side, this corresponds to including degenerate fields in the conformal block \cite{DefectAGT}. Such conformal blocks attract special interest because they satisfy natural differential equations \cite{BPZ}. This can be simply understood as a corollary of null-vector condition for the degenerate vertex operators.
\smallskip\\

Interesting phenomena occur when several generalizations from the above list are switched on at once. For example, considering together $q$-deformation and including defects leads to the problem of construction of non-stationary difference equations for $q$-conformal blocks, where certain progress has recenly been achieved in a number of cases \cite{ShiraishiNonStat,qBPZ,qPain,qBPZ2,qBPZ3}. To give another example, little is known about considering together the directions of $q$-deformation and higher genus surfaces; current paper is a modest step in this direction for genus 1 (torus). We present the summary of known results for both ordinary and $q$-deformed cases in the form of two tables: one including statements and references, another summarizing integral representations of conformal blocks.

\begin{align*}
\hspace{-5ex} \begin{array}{c|c|c|c|ccc}
\mbox{Surface} & \mbox{Vertex operators} & \mbox{Conformal block} & $q$-\mbox{deformation} & \mbox{Instanton partition function} \\ \hline
 & 4 \mbox{ generic } & \mbox{\cite{DFnew}} & \mbox{\cite{5dAGT2}} & SU(2) \mbox{ with } N_f = 4 \\
\mbox{Sphere } g = 0 & 4 \mbox{ generic and } 1 \mbox{ degenerate } & \mbox{\cite{DefectAGT}} & \mbox{\cite{qBPZ}} & SU(2)^2 \mbox{ with } N_f = 4 \mbox{ and Higgsing} \\
 & m \mbox{ generic } & \mbox{\cite{DFnew}} &  \mbox{\cite{qBPZ}} & SU(2)^{m-3} \mbox{ with } N_f = 4 \\ \hline
 & 1 \mbox{ generic } & \mbox{\cite{Toric}} & \mbox{current paper} & SU(2) \mbox{ with adjoint matter } \\
\mbox{Torus } g = 1 & 1 \mbox{ generic and }  1 \mbox{ degenerate } &  \mbox{\cite{DefectAGT}} & \mbox{current paper} &
\begin{array}{lll} SU(2) \rightleftarrows SU(2) \mbox{ with Higgsing, or } \\
 \mbox{Orbifold } SU(2) \mbox{ with adjoint matter} \end{array} \\
 & m \mbox{ generic } & \mbox{\cite{ToricGeneral}} & \mbox{current paper} & \mbox{Circular quiver of } SU(2)^{m} \\ \hline
\mbox{Higher genus } g \geq 2  & m \mbox{ generic } & \mbox{\cite{HigherGenusGeneral}} & \mbox{unknown} & \mbox{Quiver of } SU(2)^{m + 3g - 3} \mbox{ expected} \\
\end{array}
\end{align*}

{\fontsize{8pt}{0pt}{\begin{align*}
\hspace{-5ex} \begin{array}{c|c|c}
\begin{array}{c}\rule{0pt}{15mm}\end{array} \mbox{Surface} & \mbox{Conformal block} & $q$-\mbox{deformation} \\ \hline
\begin{array}{c}\rule{0pt}{15mm}\end{array}  \mbox{Sphere } & \left( \prod\limits_{i} \int dz_i \right) \prod\limits_{i \neq j} ( z_i - z_j )^{\beta} \prod\limits_{i} z_i^{a} \prod\limits_{a = 1}^{m} (x_a - z_i)^{\alpha_a} & \left( \prod\limits_{i} \int d_qz_i \right) \prod\limits_{i \neq j} \prod\limits_{m = 0}^{\beta-1} ( 1 - q^m z_i/z_j ) \prod\limits_{i} z_i^{a} \prod\limits_{a = 1}^{m} \prod\limits_{m = 0}^{\alpha_a-1} (1 - q^m z_i/x_a) \\ \hline
\begin{array}{c}\rule{0pt}{15mm}\end{array}  \mbox{Torus } & \left( \prod\limits_{i} \int dz_i \right) \prod\limits_{i \neq j} \theta_p( z_i/z_j )^{\beta} \prod\limits_{i} z_i^{a} \prod\limits_{a = 1}^{m} \theta_p(z_i/x_a)^{\alpha_a} & \left( \prod\limits_{i} \int dz_i \right) \prod\limits_{i \neq j} \prod\limits_{m = 0}^{\beta-1} \theta_p( q^m z_i/z_j ) \prod\limits_{i} z_i^{a} \prod\limits_{a = 1}^{m} \prod\limits_{m = 0}^{\alpha_a-1} \theta_p( q^m z_i/x_a) \\ \hline
\begin{array}{c}\rule{0pt}{15mm}\end{array} \mbox{Higher genus } & \mbox{Construction should be possible following \cite{HigherGenusGeneral}} & \mbox{unknown} \\
\end{array}
\end{align*}}}

One of the implications of our study is that after $q$ and elliptic deformations,
the conformal block with degenerate fields becomes \cite{ELS} the Shiraishi function,
satisfying a seemingly sophisticated equation.
What we have now is a simple integral representation of the function itself as a DF-like integral,
and the equation should have an interpretation as the null-vector condition.
Once worked out in detail, we expect this to provide both conceptual and technical clarification of this very important subject \cite{diamond}.

This paper is organized as follows. In section 2, we review the construction of Dotsenko-Fateev integral representation for multipoint conformal blocks on the sphere, including the AGT conjecture. Then in sections 3 and 4 we consider generalization to $q$-conformal blocks on the sphere, and torus conformal blocks, respectively. Finally in the main section 5 we consider both $q$-deformation and torus generalization together, presenting a $q$-deformed Dotsenko-Fateev integral for the torus case, and verifying the appropriate AGT relation. In section 6, we consider the degenerate case of the construction of section 5, and verify that it matches the Shiraishi function. Brief summary and future prospects are presented in the Conclusion section 7.

\section{The start: AGT for conformal blocks on the sphere}

As explained originally in \cite{DF} and more recently in the context of AGT conjecture in \cite{DFnew}, conformal blocks on the sphere with $m+2$ punctures can be given as multiple contour integrals of the form

\begin{align}
& \nonumber {\cal B}^{(g=0)} = \left( \int\limits_{0}^{x_1} dz_1 \ldots dz_{N_1} \right) \left( \int\limits_{0}^{x_2} dz_{N_1+1} \ldots dz_{N_1+N_2} \right) \ldots \left( \int\limits_{0}^{x_m} dz_{N_1+\ldots+N_{m-1}+1} \ldots dz_{N} \right) \\
& \prod\limits_{1 \leq i \neq j \leq N} ( z_i - z_j )^{\beta} \prod\limits_{1 \leq i \leq N} z_i^{a} (x_1 - z_i)^{\alpha_1} \ldots (x_m - z_i)^{\alpha_{m}}
\end{align}
\smallskip\\
where first group of $N_1$ variables are integrated over the segment $[0,x_1]$, next group of $N_2$ variables are integrated over the segment $[0,x_2]$, and so on until last group of $N_m$ variables integrated over the segment $[0,x_m]$. The total number of integration variables is $N = N_1 + \ldots + N_m$. Without loss of generality, we can specialize $x_1 = 1$, then conformal block is a function of $x_{2}, \ldots, x_{m}$.

An important caveat in this definition is the choice of contours. Clearly, for generic complex parameters $\beta, a, \alpha_1, \ldots, \alpha_m$ it is not viable to integrate over simple segments $[0,x_a]$ and they should be replaced by more complicated contours that respect the branching structure of the integrand. A famous example of this phenomenon is the choice of contour for the Euler beta-function: the segment $[0,1]$ is sufficient for some cases but generally it needs to be replaced by the Pochhammer contour. To avoid complexities related to the choice of contours, in \cite{DFnew} we suggested to use the simplest variant of analytic continuation for ${\cal B}^{(g=0)}$: declare all parameters $\beta, a, \alpha_1, \ldots, \alpha_m$ as positive integers, and all contours as simple segments. This is sufficient for many purposes: indeed, if ${\cal B}^{(g=0)}$ is represented as a power series in $x_{2}, \ldots, x_{m}$, then every term in this series is a rational function of $\beta, a, \alpha_1, \ldots, \alpha_m, N_1, \ldots, N_m$ and is fully determined by the values at positive integer locus.

To give an example, for $m = 2$ this gives integral representation of the 4-point spherical conformal block,

\begin{align}
{\mathcal B}^{(g=0,m=2)}(x) = \int\limits_{0}^{1} dz_{1} \ldots dz_{N_1} \int\limits_{0}^{x} dz_{N_1 + 1} \ldots dz_{N_1 + N_2} \ \prod\limits_{i \neq j} (z_i - z_j)^{\beta} \prod\limits_{i} z_1^a (1 - z_i)^{\alpha} (x - z_i)^{\alpha_2}
\end{align}
\smallskip\\
It is easy to verify in series that

\begin{align}
{\mathcal B}^{(g=0,m=2)}(x) = \mbox{const} \cdot x^{d} (1-x)^{\frac{-\alpha \alpha_2}{2 \beta}} \big( 1 + B_1 x + B_2 x^2 + \ldots \big)
\label{Bseries}
\end{align}
\smallskip\\
where $d = N_2 + \alpha_2 N_2 + a N_2 + \beta N_2(N_2-1)$ and

\begin{align}
B_1 = \dfrac{(\Delta + \Delta_2 - \Delta_1) (\Delta + \Delta_3 - \Delta_4)}{2 \Delta}
\end{align}
\begin{align}
\nonumber B_2 = \dfrac{(\Delta + \Delta_2 - \Delta_1) (\Delta + \Delta_2 - \Delta_1+1)(\Delta + \Delta_3 - \Delta_4)(\Delta + \Delta_3 - \Delta_4 + 1)}{4 \Delta( 2 \Delta + 1)} + \\
\dfrac{((\Delta_1 + \Delta_2) (2 \Delta + 1) + \Delta (\Delta-1) - 3 (\Delta_1 - \Delta_2)^2)((\Delta_3 + \Delta_4) (2 \Delta + 1) + \Delta (\Delta-1) - 3 (\Delta_3 - \Delta_4)^2)}{2(2\Delta+1)(2\Delta(8\Delta - 5)+(2\Delta + 1) c)}
\end{align}
\smallskip\\
with

\begin{align}
& \Delta_1 = \dfrac{ a (a + 2 - 2 \beta)}{ 4 \beta }, \ \ \ \Delta_2 = \dfrac{ \alpha_2 (\alpha_2 + 2 - 2 \beta)}{4 \beta}, \ \ \ \Delta_3 = \dfrac{ \alpha (\alpha + 2 - 2 \beta)}{ 4 \beta }, \\
& \Delta_4 = \dfrac{ (a + \alpha + \alpha_2 + 2 \beta (N_1+N_2)) (a + \alpha + \alpha_2 + 2 \beta (N_1+N_2) + 2 - 2 \beta)}{ 4 \beta }, \\
& \Delta = \dfrac{ (a + \alpha_2 + 2 \beta N_2)(a + \alpha_2 + 2 \beta N_2 + 2 - 2 \beta)}{ 4 \beta }, \\
& c = 1 - 6 (\beta - 2 + 1/\beta)
\end{align}
\smallskip\\
One can see that particular terms in the series such as $B_1, B_2, \ldots$ are indeed rational functions as we discussed before. The coefficients in the series expansion $B_1,B_2,\ldots$ match precisely the well-known formulas for 4-point conformal blocks on the sphere, obtained by the OPE methods \cite{OPE}.

\begin{wrapfigure}{l}{0.55\textwidth}
{\vspace{-6ex}
  \begin{center}
    \includegraphics[width=0.55\textwidth]{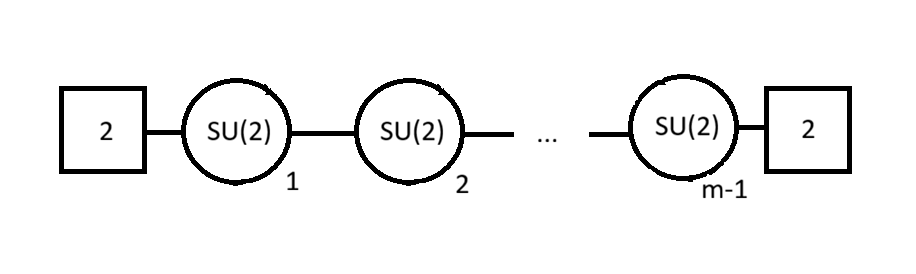}
  \end{center}
  \vspace{-6ex}}
\caption{Linear quiver of $m-1$ gauge groups of type $SU(2)$. The theory has 4 fundamental matter multiplets shown in rectangular boxes.}
\vspace{1ex}
\end{wrapfigure} \paragraph{}On the gauge theory side, this series is known \cite{OPE} to match the instanton partition function of four-dimensional $SU(2)$ theory with 4 fundamental matter multiplets. More generally, for $m \geq 2$, conformal block is known \cite{OPEhigher} to match the instanton partition function of four-dimensional $SU(2)^{m-1}$ theory with linear quiver, and 4 fundamentals, as shown on Figure 2. This summarizes the AGT relation for the case of the sphere, and will be our starting point for generalization.

\section{$q$-deformation on the sphere}

The first generalization we consider is $q$-deformation. Here, $q$-deformation of Virasoro algebra \cite{q-Virasoro} gives rise to a natural $q$-deformation of spherical conformal blocks \cite{5dAGT2} (see \cite{qBPZ} for a recent exposition)

\begin{align}
& \nonumber {\cal B}_q^{(g=0)} = \left( \int\limits_{0}^{x_1} d_qz_1 \ldots d_qz_{N_1} \right) \left( \int\limits_{0}^{x_2} d_qz_{N_1+1} \ldots d_qz_{N_1+N_2} \right) \ldots \left( \int\limits_{0}^{x_m} d_qz_{N_1+\ldots+N_{m-1}+1} \ldots d_qz_{N} \right) \\
& \prod\limits_{1 \leq i \neq j \leq N} \prod\limits_{k = 0}^{\beta-1} ( 1 - q^k z_i/z_j ) \prod\limits_{1 \leq i \leq N} z_i^{a} \prod\limits_{k = 0}^{\alpha_1 - 1} (1 - q^k z_i/x_1) \ldots \prod\limits_{k = 0}^{\alpha_m - 1} (1 - q^k z_i/x_m)
\end{align}
\smallskip\\
In simple terms, this deformation has two important effects: first, at the level of two-point functions the powers of differences\footnote{At this point, to simplify the exposition we trade $(y-x)^{\alpha}$ for $(1-x/y)^{\alpha}$ at the cost of simple relation of parameters.} are simply replaced by their $q$-shifted counterparts,

\begin{align}
(1-x)^{\alpha} \ \ \ \rightarrow \ \ \ (1-x) (1 - qx) \ldots (1 - q^{\alpha-1} x)
\end{align}
\smallskip\\
and second, the contour integrals are replaced by discretized $q$-integrals so that

\begin{align}
\int\limits_{0}^{x} dz \ z^{n} = \dfrac{x^{n+1}}{n+1} \ \ \ \rightarrow \ \ \ \int\limits_{0}^{x} d_qz \ z^{n} = \dfrac{(1-q) x^{n+1}}{1-q^{n+1}}
\end{align}
\smallskip\\
Again, we assume that $\beta, a, \alpha_1, \ldots, \alpha_m$ are positive integers. Then, ${\cal B}_q^{(g=0)}$ can be represented as a power series in $x_2, \ldots, x_m$ where every term is a rational function in $q^\beta, q^a, q^{\alpha_1}, \ldots, q^{\alpha_m}, q^{\beta N_1}, \ldots, q^{\beta N_m}$. Therefore the positive integer locus and the above simple choice of contours ($q$-integrals) is sufficient to describe this series completely. On the gauge theory side, this series is known for $m=2$ \cite{5dAGT2} to match the instanton partition function of five-dimensional $SU(2)$ theory with 4 fundamental matter multiplets. For $m > 2$, it is expected to match the instanton partition function of five-dimensional $SU(2)^{m-1}$ theory with linear quiver, and 4 fundamentals.

\section{AGT for conformal blocks on the torus}

Next generalization we consider is conformal blocks on a torus. Here, as explained in \cite{ToricGeneral} and further developed in \cite{Toric}, to construct a contour integral representation of a torus conformal block it suffices to replace at the level of two-point functions

\begin{align}
(1-x)^{\alpha} \ \ \ \rightarrow \ \ \ \theta_p(x)^{\alpha}, \ \ \ \theta_p(x) := \prod\limits_{n = 0}^{\infty} (1 - p^n x)(1 - p^{n+1}/x)
\end{align}
\smallskip\\
and do not discard an important factor $\prod_i z_i^a$. On the sphere this factor used to represent the two-point function with the insertion at 0 but on the torus it has a completely different meaning of exponentiated integral of a holomorphic differential. This results in the following integral for the $m$-point conformal block on the torus,

\begin{align}
& \nonumber {\cal B}^{(g=1)} = \left( \int\limits_{0}^{x_1} d_qz_1 \ldots d_qz_{N_1} \right) \left( \int\limits_{0}^{x_2} d_qz_{N_1+1} \ldots d_qz_{N_1+N_2} \right) \ldots \left( \int\limits_{0}^{x_m} d_qz_{N_1+\ldots+N_{m-1}+1} \ldots d_qz_{N} \right) \\
& \prod\limits_{1 \leq i \neq j \leq N} \theta_p( z_i/z_j )^{\beta} \prod\limits_{1 \leq i \leq N} z_i^{a} \theta_p(z_i/x_1)^{\alpha_1} \ldots \theta_p(z_i/x_m)^{\alpha_m}
\end{align}
\smallskip\\
where one condition $\alpha_1 + \ldots + \alpha_m + 2 \beta N = 0$ should be imposed for charge conservation. Applying the same method as before, we assume $\beta, a, \alpha_1, \ldots, \alpha_m$ are positive integers, specialize $x_1 = 1$ without loss of generality, and expand ${\cal B}^{(g=1)}$ into power series in $p, x_2, \ldots, x_m$. Then, each term in the series expansion is a rational function of parameters and can be completely determined from the positive integer locus. On the gauge theory side, this series is known for $m=1$ \cite{AGTToric} to match the instanton partition function of four-dimensional $SU(2)$ theory with adjoint matter. More generally, for $m > 1$, the correspondence to four-dimensional $SU(2)^{m}$ theory with circular quiver was established in \cite{ToricGeneral}.

\section{$q$-deformation on the torus}

It is interesting to see, how $q$-deformation interferes with generalization to a torus. Already the question of deformation for the two-point function is somewhat non-trivial. We suggest the following option

\begin{align}
\begin{array}{ccccccc}
& & (1-x) (1 - qx) \ldots (1 - q^{\alpha-1} x) & & \\
& \nearrow & & \searrow & \\
(1-x)^{\alpha} & & & & \theta_p(x) \theta_p(qx) \ldots\theta_p(q^{\alpha-1} x) \\
& \searrow &  & \nearrow & \\
& & \theta_p(x)^{\alpha} & &
\end{array}
\end{align}
\smallskip\\
and justify its validity with several checks below. Of course, it would be desirable to relate this choice of two-point functions to representation theory of $q$-Virasoro algebra. With this choice, we have the following $q$-deformation of the $m$-point conformal block on the torus,

\begin{align}
& \nonumber {\cal B}_q^{(g=1)} = \left( \int\limits_{0}^{x_1} d_qz_1 \ldots d_qz_{N_1} \right) \left( \int\limits_{0}^{x_2} d_qz_{N_1+1} \ldots d_qz_{N_1+N_2} \right) \ldots \left( \int\limits_{0}^{x_m} d_qz_{N_1+\ldots+N_{m-1}+1} \ldots d_qz_{N} \right) \\
& \prod\limits_{1 \leq i \neq j \leq N} \prod\limits_{k = 0}^{\beta-1} \theta_p( q^k z_i/z_j ) \prod\limits_{1 \leq i \leq N} z_i^{a} \prod\limits_{k = 0}^{\alpha_1 - 1} \theta_p(q^k z_i/x_1) \ldots \prod\limits_{k = 0}^{\alpha_m - 1} \theta_p(q^k z_i/x_m)
\label{qBlockTorus}
\end{align}
\smallskip\\
where as before, we impose the condition $\alpha_1 + \ldots + \alpha_m + 2 \beta N = 0$ for charge conservation. The method that we used before can be applied again: assuming $\beta, a, \alpha_1, \ldots, \alpha_m$ are positive integers, specializing $x_1 = 1$ and expanding ${\cal B}_q^{(g=1)}$ into power series in $p, x_2, \ldots, x_m$. Then, each term in the series expansion is a rational function of $q^\beta, q^a, q^{\alpha_1}, \ldots, q^{\alpha_m}, q^{\beta N_1}, \ldots, q^{\beta N_m}$ and can be completely determined from the positive integers.

On the gauge theory side, we expect ${\cal B}_q^{(g=1)}$ to match the instanton partition function of five-dimensional $SU(2)^{m}$ theory with circular quiver. This partition function can be given as a sum over $2 m$ partitions (Young diagrams) $Y_{a,i}$ with $a = 1, \ldots, m$ and $i = 1, 2$:

{\fontsize{9pt}{0pt}{\begin{align}
Z = \mathop{\mathop{\sum\limits_{\{ Y_{a,i} \}}}_{1 \leq a \leq m }}_{1 \leq i \leq 2} \ \prod\limits_{i,j = 1}^{2} \dfrac{{\cal N}_{Y_{1,i}, Y_{2,j}}\big( f_{1} Q_{1,i}/Q_{2,j} \big) {\cal N}_{Y_{2,i}, Y_{3,j}}\big( f_{2} Q_{2,i}/Q_{3,j} \big) \ldots {\cal N}_{Y_{m,i}, Y_{1,j}}\big( f_{m} Q_{m,i}/Q_{1,j} \big) }{ {\cal N}_{Y_{1,i}, Y_{1,j}}\big( Q_{1,i}/Q_{1,j} \big) \ldots {\cal N}_{Y_{m,i}, Y_{m,j}}\big( Q_{m,i}/Q_{m,j} \big) } \prod\limits_{a = 1}^{m} \Lambda_a^{|Y_{a,1}| + |Y_{a,2}|}
\label{qCircularTheory}
\end{align}}}
\smallskip\\
We remind that a partition is a sequence of the form $\lambda = (\lambda_1 \geq \lambda_2 \geq \ldots \geq \lambda_{\ell(\lambda)} \geq 0)$. The summand is a product of 5d Nekrasov factors, defined for a pair of partitions $A,B$ as

\begin{align}
{\cal N}_{A,B}(Q) = \prod\limits_{(i,j) \in A} (1 - Q q^{A_i - j} t^{B^T_j - i + 1}) \prod\limits_{(i,j) \in B} (1 - Q q^{- B_i + j - 1} t^{- A^T_j + i})
\end{align}
\smallskip\\
where $(i,j) \in A$ is a shorthand for $i = 1, \ldots, \ell(A)$ and $j = 1, \ldots, A_i$. The length of $A$ (number of parts in the partition) is denoted $\ell(A)$ and the size of $A$ (the sum of all parts) is denoted $|A|$. The partition function depends on $m$ parameters $f_1, \ldots, f_{m}$ (exponentiated masses of bifundamentals) and $2m$ parameters $Q_{a,i}$ with $a = 1, \ldots, m$ and $i = 1, 2$ (exponentiated Coulomb parameters) and is a series in $m$ variables $\Lambda_1, \ldots, \Lambda_m$ (the instanton counting parameters).

We will perform comparison between ${\cal B}_q^{(g=1)}$ and $Z$ for the minimal non-trivial value of $m=2$. For this case, the $q$-deformed conformal block can be written as

\begin{align}
& \nonumber {\cal B}_q^{(g=1,m=2)}(p,x) = \int\limits_{0}^{1} d_qz_1 \ldots d_qz_{N_1} \int\limits_{0}^{x} d_qz_{N_1+1} \ldots d_qz_{N_1+N_2} \\
& \prod\limits_{1 \leq i \neq j \leq N} \prod\limits_{k = 0}^{\beta-1} \theta_p( q^k z_i/z_j ) \prod\limits_{1 \leq i \leq N} z_i^{a} \prod\limits_{k = 0}^{\alpha_1 - 1} \theta_p(q^k z_i) \prod\limits_{k = 0}^{\alpha_2 - 1} \theta_p(q^k z_i/x)
\end{align}
\smallskip\\
where $N = N_1 + N_2$ and charge conservation $\alpha_1 = - 2 \beta N - \alpha_2$ is assumed. It is easy to verify in series that

\begin{align}
{\cal B}_q^{(g=1,m=2)}(p,x) = \mbox{const} \cdot x^{-\alpha_2 N_1 + \beta N_2 (N_2-1)} \sum\limits_{i,j = 0}^{\infty} {\cal B}_{i,j} x^i (p/x)^j
\label{Bseries2}
\end{align}
\smallskip\\
Direct calculation gives a first few coefficients: ${\cal B}_{0,0} = 1$ and

\begin{center}
$
{\cal B}_{1,0} = \dfrac{q}{X X_2^2 X_1^2 (q - 1) (\tau - 1) (A X X_2^2 q - \tau^2) (A X X_2^2 - q)} \cdot \Big( A^2 X^2 X_1^2 X_2^6 \tau - A^2 X^2 X_1^2 X_2^5 \tau - A X^2 X_1^2 X_2^5 \tau^2 + A X^2 X_1^2 X_2^4 q \tau + A X^2 X_1^2 X_2^4 \tau^2 - A^2 X^2 X_1 X_2^4 q - A X^2 X_1 X_2^4 q \tau - A X X_1^2 X_2^4 q \tau + A X X_1^2 X_2^4 \tau^2 + A^2 X^2 X_2^4 q - A^2 X X_1 X_2^4 \tau - A X X_1^2 X_2^3 \tau^2 - A X X_1 X_2^4 \tau^2 - X X_1^2 X_2^3 \tau^3 + A^2 X X_1 X_2^3 q + A^2 X X_1 X_2^3 \tau + 2 A X X_1 X_2^3 q \tau + 2 A X X_1 X_2^3 \tau^2 + X X_1 X_2^3 q \tau^2 + X X_1 X_2^3 \tau^3 - A^2 X X_2^3 q - A X X_1 X_2^2 q \tau - A X X_2^3 q \tau - X X_1 X_2^2 q \tau^2 + X_1^2 X_2^2 \tau^3 + A X X_2^2 q \tau - A X X_2^2 \tau^2 - A X_1 X_2^2 \tau^2 - X_1 X_2^2 \tau^3 + A X_2^2 q \tau + A X_2^2 \tau^2 - A X_2 q \tau - X_2 q \tau^2 + q \tau^2 \Big)
$
\end{center}

\begin{center}
$
{\cal B}_{0,1} = \dfrac{q}{X_2 (q - 1) (\tau - 1) (-q + A) (A q - \tau^2)} \cdot \Big( A^2 X^2 X_1^2 X_2^4 \tau + A X^2 X_1^2 X_2^4 \tau^2 - A^2 X^2 X_1 X_2^4 q - A^2 X^2 X_1 X_2^4 \tau - A X^2 X_1^2 X_2^3 q \tau - A X^2 X_1^2 X_2^3 \tau^2 + A^2 X^2 X_1 X_2^3 q - A^2 X X_1^2 X_2^3 \tau + A X^2 X_1 X_2^3 q \tau + A X X_1^2 X_2^3 q \tau - A X X_1^2 X_2^3 \tau^2 - X X_1^2 X_2^3 \tau^3 + A^2 X X_1 X_2^3 \tau + A X X_1^2 X_2^2 \tau^2 + A X X_1 X_2^3 \tau^2 + X X_1^2 X_2^2 \tau^3 - 2 A X X_1 X_2^2 q \tau - 2 A X X_1 X_2^2 \tau^2 + A^2 X X_2^2 q + A X X_1 X_2 q \tau + A X X_2^2 q \tau + X X_1 X_2 q \tau^2 - A^2 X X_2 q - A X X_2 q \tau + A X X_2 \tau^2 + A X_1 X_2 \tau^2 - X X_2 q \tau^2 + X_1 X_2 \tau^3 - A X_2 q \tau - A X_2 \tau^2 - X_1 q \tau^2 - X_1 \tau^3 + A q \tau + q \tau^2 \Big)
$
\end{center}
where we denoted $\tau = q^{\beta}, \ A = q^{a+1}, \ X_1 = q^{\beta N_1}, \ X_2 = q^{\beta N_2}, \ X = q^{\alpha_2}$.

At the same time, the instanton partition function for $m=2$ takes form

\begin{align}
Z^{(m=2)} = \sum\limits_{ Y_{1,1}, Y_{1,2}, Y_{2,1}, Y_{2,2} }\ \prod\limits_{i,j = 1}^{2} \dfrac{{\cal N}_{Y_{1,i}, Y_{2,j}}\big( f_{1} g_i/h_j \big) {\cal N}_{Y_{2,i}, Y_{1,j}}\big( f_{2} h_i/g_j \big) }{ {\cal N}_{Y_{1,i}, Y_{1,j}}\big( g_i/g_j \big) {\cal N}_{Y_{2,i}, Y_{2,j}}\big( h_i/h_j \big) } \ \Lambda_1^{|Y_{1,1}| + |Y_{1,2}|} \Lambda_2^{|Y_{2,1}| + |Y_{2,2}|}
\end{align}
\smallskip\\
where we denoted $Q_{1,i} = g_i, Q_{2,i} = h_i$. It is easy to verify in series that

\begin{align}
Z^{(m=2)} = \sum\limits_{i,j = 0}^{\infty} {\cal Z}_{i,j} \Lambda_1^i \Lambda_2^j
\end{align}
\smallskip\\
Direct calculation gives a first few coefficients: ${\cal Z}_{0,0} = 1$ and

\begin{align*}
{\cal Z}_{1,0} = \dfrac{(1 - \frac{f_1 g_2}{h_1}) (1 - \frac{ f_1 g_2 }{h_2} ) (1 - \frac{f_2 h_1 t}{g_2 q}) (1 - \frac{f_2 h_2 t}{g_2 q})}{ (1 - \frac{g_1 t}{g_2 q}) (1 - \frac{g_2}{g_1}) (1 - t) (1 - q^{-1})} + \dfrac{(1 - \frac{f_1 g_1}{h_1}) (1 - \frac{f_1 g_1}{h_2}) (1 - \frac{f_2 h_1 t}{ g_1 q }) (1 - \frac{f_2 h_2 t}{ g_1 q })}{(1 - t)(1 - q^{-1}) (1 - \frac{g_1}{g_2}) (1 - \frac{ g_2 t}{ g_1 q })}
\end{align*}
\begin{align*}
{\cal Z}_{0,1} = \dfrac{(1 - \frac{f_1 g_1 t}{h_2 q}) (1 - \frac{f_1 g_2 t}{h_2 q}) (1 - \frac{f_2 h_2}{g_1}) (1 - \frac{f_2 h_2}{g_2})}{(1 - \frac{h_1 t}{h_2 q}) (1 - \frac{h_2}{h_1}) (1 - t) (1 - q^{-1})} + \dfrac{(1 - \frac{f_1 g_1 t}{h_1 q}) (1 - \frac{f_1 g_2 t}{h_1 q}) (1 - \frac{f_2 h_1}{g_1}) (1 - \frac{f_2 h_1}{g_2})}{(1 - t) (1 - q^{-1}) (1 - \frac{h_1}{h_2}) (1 - \frac{h_2 t}{h_1 q})}
\end{align*}
\smallskip\\
This is just an example. We actually computed coefficients ${\cal B}_{i,j}$ and ${\cal Z}_{i,j}$ up to level 4 (i.e., $0 \leq i+j \leq 4$). Based on this information, we propose the following exact relation between the 5d instanton partition function and the $q$-deformed conformal block:

\begin{align}
\boxed{ \ \ \ \sum\limits_{i,j = 0}^{\infty} {\cal Z}_{i,j} \Lambda_1^i \Lambda_2^j = \sigma(p,x) \sum\limits_{i,j = 0}^{\infty} {\cal B}_{i,j} x^i (p/x)^j \ \ \ }
\label{5dTorusAGT}
\end{align}
\smallskip\\
under identification of parameters

\begin{align}
\boxed{ \ \ \ \begin{array}{ccc} \Lambda_1 = q^{\beta-1} x, \ \ \ \Lambda_2 = q^{\beta-1+2\alpha_2+4 \beta N} p/x, \ \ \ t = q^{\beta} \\
\\
h_1/h_2 = q^{a+1-\beta}, \ \ \ g_1/g_2 = q^{a+\alpha_2+2\beta N_2 +1-\beta} \\
\\
f_1 f_2 = q^{1-\beta+\alpha_2+\beta N}, \ \ \ f_2 h_2 / g_2 = q^{1 - \beta - \beta N_1}
 \end{array} \ \ \ }
\label{IdentificationCircular}
\end{align}
\smallskip\\
where the correction factor $\sigma(p,x)$ is given by

\begin{align}
\boxed{ \ \begin{array}{lll} \sigma(p,x) = & \mbox{pexp}\left( \dfrac{p}{x} \ \dfrac{q^{\alpha_2-1}(t-q)(t^{2N + 2} - q^2)}{(1-q)(1-t)(1-p)} \right) \times \\ & \\ & \mbox{pexp}\left( - p \ \dfrac{N (q+t)(1-t)^2 + q^{\alpha_2} t^{2 N + 1} + q^{- \alpha_2 + 1} t^{- 2 N} + t q^{-\alpha_2} + q^{\alpha_2 + 1} + 2 (1 + t q) }{(1-q)(1-t)(1-p)} \right) \end{array} \ }
\end{align}
\smallskip\\
We remind that plethystic exponent of any rational function $f(q,t,p,x)$ is defined as

\begin{align}
\mbox{pexp}\left( f(q,t,p,x) \right) = \exp\left( \sum\limits_{k = 1}^{\infty} \dfrac{f(q^k,t^k,p^k,x^k)}{k} \right)
\end{align}
\smallskip\\
Remarkably, the correction factor $\sigma(p,x)$ is completely independent of $a$ and depends only on $N_1+N_2$, not on $N_1$ and $N_2$ separately. Therefore, we verified up to level 4 that $q$-deformed torus conformal block \begin{wrapfigure}{r}{0.35\textwidth}
{
  \begin{center}
    \includegraphics[width=0.35\textwidth]{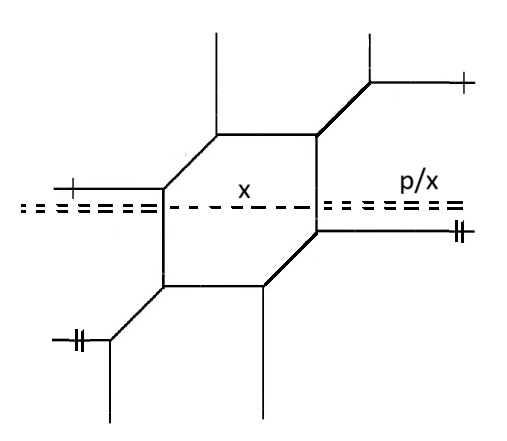}
  \end{center}
  }
\caption{Topological string geometry that engineers the gauge theory with circular quiver $SU(2)^2$.}
\vspace{-5ex}
\end{wrapfigure}matches the 5d instanton partition function of a circular quiver, up to relatively simple $a$-independent prefactor. This is an interesting check of AGT conjecture for $q$-deformed theories and beyond the sphere. Generally, for $m \geq 2$ we expect that the block (\ref{qBlockTorus}) matches the partition function (\ref{qCircularTheory}) up to $a$-independent prefactor.

A short note is in order: the prefactor's dependence on $x$ is fully captured by a simple plethystic
exponent

\begin{align*}
\mbox{pexp}\left( \dfrac{p}{x} \ \dfrac{q^{\alpha_2-1}(t-q)(t^{2N + 2} - q^2)}{(1-q)(1-t)(1-p)} \right)
\end{align*}
\smallskip\\

We note that this simple factor is reminiscent of contributions of non-compact BPS states associated with infinite parallel vertical lines in the topological string toric geometry that engineers the gauge theory with circular quiver $SU(2)^2$, as shown on Figure 3. This may shed some light on the nature of function $\sigma(p,x)$. The second factor in $\sigma(p,x)$ does not depend on $x$ and is just a $p$-dependent constant.

\section{Defects for a circular quiver: Shiraishi function}

Having considered a generic case of 5d AGT correspondence with a circular quiver, let us turn attention to introducing defects. As we already mentioned in the Introduction, on the conformal field theory side this corresponds to degenerate fields or vertex operators. On the gauge theory side of AGT correspondence, we expect to see the partition function of a circular quiver with defect, also known as the Shiraishi function \cite{ShiraishiNonStat}.

It is well-known that to degenerate the conformal block one needs to introduce two constraints: see, for example, \cite{DefectAGT} or more recent \cite{qBPZ}. In the Dotsenko-Fateev formalism, these are constraints on parameters $\alpha_a$ and $N_a$ associated with the degenerate insertion point $z_a$. Instead of going into details of derivation of this fact from first principles of conformal field theory, let us simply fix these constraints in our current notation from a computation. It is instructive to consider again the simplest case of 4-point conformal block on the sphere,

\begin{align}
{\mathcal B}^{(g=0,m=2)}(x) = \int\limits_{0}^{1} dz_{1} \ldots dz_{N_1} \int\limits_{0}^{x} dz_{N_1 + 1} \ldots dz_{N_1 + N_2} \ \prod\limits_{i \neq j} (z_i - z_j)^{\beta} \prod\limits_{i} z_1^a (1 - z_i)^{\alpha} (x - z_i)^{\alpha_2}
\end{align}
\smallskip\\
and look for values of $(\alpha_2, N_2)$ where the function ${\mathcal B}^{(g=0,m=2)}(x)$ satisfies a hypergeometric second-order differential equation. Direct computation with the series (\ref{Bseries}) shows that there are 4 possible solutions $(\alpha_2, N_2)$:

\begin{align}
& \mbox{A)} N_2 = 0, \ \ \ \alpha_2 = +1 \\
& \mbox{B)} N_2 = 1, \ \ \ \alpha_2 = -\beta \\
& \mbox{C)} N_2 = 0, \ \ \ \alpha_2 = -\beta \\
& \mbox{D)} N_2 = - \beta^{-1}, \ \ \ \alpha_2 = +1
\end{align}
\smallskip\\
Note that A) was the preferred choice of \cite{DefectAGT} while B) was the preferred choice of \cite{qBPZ}. The corresponding DF integrals are:

\begin{align}
& {\mathcal B}^{(A)}(x) = \int\limits_{0}^{1} dz_{1} \ldots dz_{N_1} \ \prod\limits_{i \neq j} (z_i - z_j)^{\beta} \prod\limits_{i} z_1^a (1 - z_i)^{\alpha} (x - z_i) \\
& {\mathcal B}^{(B)}(x) = \int\limits_{0}^{1} dz_{1} \ldots dz_{N_1} \int\limits_{0}^{w} dz_{N_1 + 1} \ \prod\limits_{i \neq j} (z_i - z_j)^{\beta} \prod\limits_{i} z_1^a (1 - z_i)^{\alpha} (x - z_i)^{-\beta} \\
& {\mathcal B}^{(C)}(x) = \int\limits_{0}^{1} dz_{1} \ldots dz_{N_1} \ \prod\limits_{i \neq j} (z_i - z_j)^{\beta} \prod\limits_{i} z_1^a (1 - z_i)^{\alpha} (x - z_i)^{-\beta} \\
& {\mathcal B}^{(D)}(x) = \mbox{analytical continuation of } {\mathcal B}^{(g=0,m=2)}(x) \mbox{ to } N_2 = -1/\beta \mbox{ with } \alpha_2 = + 1
\end{align}
\smallskip\\
Direct series evaluation gives

\begin{align}
& {\mathcal B}^{(A)}(x) \ = \ \mbox{const} \cdot {}_{2}F_{1} \left( \begin{array}{cc} \frac{N_1 \beta+a+\alpha+2-\beta}{\beta} & - N_1 \\ \frac{1+a}{\beta} \end{array} \Big| \ x \ \right) \\
& {\mathcal B}^{(B)}(x) \ = \ \mbox{const} \cdot x^{a - \beta + 1} {}_{2}F_{1} \left( \begin{array}{cc} N \beta - \beta + 1 + a & \beta - N \beta - \alpha \\ a + 2 - \beta \end{array} \Big| \ x \ \right) \\
& {\mathcal B}^{(C)}(x) \ = \ \mbox{const} \cdot {}_{2}F_{1} \left( \begin{array}{cc} - a - \alpha - \beta N_1 + 2 \beta - 1 & \beta N_1 \\ \beta - a \end{array} \Big| \ x \ \right) \\
& {\mathcal B}^{(D)}(x) \ = \ \mbox{const} \cdot x^{1-\frac{1+a}{\beta}} {}_{2}F_{1} \left( \begin{array}{cc} N_1 + \frac{\alpha}{\beta} & 1 - N_1 - \frac{a}{\beta} \\ 2-\frac{1+a}{\beta} \end{array} \Big| \ x \ \right)
\end{align}
\smallskip\\
At this point we can comment on existence of second solution for a hypergeometric differential equation of second order. Indeed, it is well-known that pair of hypergeometric functions

\begin{align*}
 {}_{2}F_{1} \left( \begin{array}{cc} a & b \\ c \end{array} \Big| \ x \ \right), \ \ \ x^{1-c} {}_{2}F_{1} \left( \begin{array}{cc} 1+a-c & 1+b-c \\ 2-c \end{array} \Big| \ x \ \right)
\end{align*}
\smallskip\\
give two solutions to the same hypergeometric equation. Looking at the formulas above, we conclude that ${\mathcal B}^{(B)}(x), {\mathcal B}^{(C)}(x)$ are such and ${\mathcal B}^{(A)}(x), {\mathcal B}^{(D)}(x)$ are such.

Therefore, two complementary solutions to the same differential equation are always associated with two different choices of contour for the same null-vector and same potential (parameter $\alpha_2$) of the matrix model. However choice of contour is intricate: while for the pair ${\mathcal B}^{(B)}(x), {\mathcal B}^{(C)}(x)$ it is simple: 1 integral around $[0,1]$ vs. 1 integral around $[0,x]$, for the other pair ${\mathcal B}^{(A)}(x), {\mathcal B}^{(D)}(x)$ we need 1 integral around $[0,1]$ vs. $-1/\beta$ integrals around $[0,x]$. This is probably the reason the branch D) was overlooked in \cite{DefectAGT}.

Having fixed the pair of constraints necessary to consider a degenerate conformal block, let us now apply this to our case of interest: $q$-deformed conformal blocks on the torus. As mentioned above, we proceed simply by generalizing the hypergeometric cases A)-D). For brevity, we will only consider one option A) out of four. The result is a $q$-deformed integral

\begin{align}
{\cal B}_q^{(A)}(p,x) = \int\limits_{0}^{1} d_qz_1 \ldots d_qz_{N} \prod\limits_{1 \leq i \neq j \leq N} \prod\limits_{k = 0}^{\beta-1} \theta_p( q^k z_i/z_j ) \prod\limits_{1 \leq i \leq N} z_i^{a} \theta_p(q^k z_i/x) \prod\limits_{k = 0}^{\alpha_1 - 1} \theta_p(q^k z_i)
\end{align}
\smallskip\\
where condition $\alpha_1 = -1 - 2 \beta N$ needs to be imposed after evaluation. Function ${\cal B}_q^{(A)}(p,x)$ will be our primary candidate to match the Shiraishi function. It is easy to verify in series that

\begin{align}
{\cal B}_q^{(A)}(p,x) = \mbox{const} \cdot x^{-N_1} \sum\limits_{i,j = 0}^{\infty} {\cal B}^{(A)}_{i,j} x^i (p/x)^j
\label{Bseries3}
\end{align}
\smallskip\\
Direct calculation gives a first few coefficients: ${\cal B}^{(A)}_{0,0} = 1$ and

\begin{align}
& {\cal B}^{(A)}_{1,0} = \dfrac{- q^{\beta - 1 - \beta N} (1 - q^{\beta N})(1 - q^{a - \beta N + 1 - \beta})}{(1 - q^{\beta})(1 - q^{a+1})}, \\
\nonumber \\
& {\cal B}^{(A)}_{0,1} = \dfrac{- q (1 - q^{\beta N})(1 - q^{a + \beta N + 2 - \beta})}{(1 - q^{\beta})(1 - q^{a+2 - 2 \beta})}
\end{align}
\smallskip\\
At the same time, Shiraishi function of rank 1 is defined as a sum over two partitions,

\begin{align}
Z^{\mbox{(defect)}}\big( x_1, x_2, p \big\vert y_1, y_2, s \big) = \sum\limits_{\lambda^{(1)}, \lambda^{(2)}} \ \prod\limits_{i,j = 1}^{2} \frac{{\cal N}_{\lambda^{(i)}, \lambda^{(j)}}^{(j-i)}\big( {\widetilde q}/{\widetilde t} \ y_j / y_i \vert {\widetilde q}, s \big)}{{\cal N}_{\lambda^{(i)}, \lambda^{(j)}}^{(j-i)}\big( y_j / y_i \vert {\widetilde q}, s \big)} \ \prod\limits_{b = 1}^{2} \prod\limits_{a \geq 1} \left( \frac{p {\widetilde t} x_{a+b}}{{\widetilde q} x_{a+b-1}} \right)^{\lambda^{(b)}_a}
\end{align}
\smallskip\\
where the summand is a product of orbifold Nekrasov factors

\begin{align}
& {\cal N}_{\lambda, \mu}^{(k)}\big( u \vert {\widetilde q}, s \big) =
\prod\limits_{1 \leq a \leq b \leq \ell(\lambda)} \prod\limits_{m = 0}^{\lambda_b - \lambda_{b+1} - 1} \big( 1 - u {\widetilde q}^m {\widetilde q}^{-\mu_a + \lambda_{b+1}} s^{-a+b} \big)^{\delta_{b - a - k \vert N}} \\
& \prod\limits_{1 \leq a \leq b \leq \ell(\mu)} \prod\limits_{m = 0}^{\mu_b - \mu_{b+1} - 1}
\big( 1 - u {\widetilde q}^m {\widetilde q}^{\lambda_{a} - \mu_b} s^{a-b-1} \big)^{\delta_{b - a + k + 1 \vert N}}
\end{align}
\smallskip\\
with $\delta_{n \vert N} = 1$ if $n \bmod N = 0$ and $\delta_{n \vert N} = 0$ otherwise. Here $x_{n + 2} = x_{n}$. For convenience, we also define

\begin{align}
\psi(X,Y|{\widetilde q},{\widetilde t}|p,s) = Z^{\mbox{(defect)}}\big( X^{1/2} p^{-1/2}, X^{-1/2}, p^{1/2} \big\vert Y^{1/2} s^{-1/2}, Y^{-1/2}, s^{1/2} \big)
\end{align}
\smallskip\\
It is easy to verify in series that

\begin{align}
\psi(X,Y|{\widetilde q},{\widetilde t}|p,s) = \sum\limits_{i,j = 0}^{\infty} \psi_{i,j} X^i (p/X)^j
\end{align}
\smallskip\\
Direct calculation gives a first few coefficients: $\psi_{0,0} = 1$ and

\begin{align}
& \psi_{1,0} = \dfrac{{\widetilde t} (1 - \frac{1}{{\widetilde t}}) (1 - \frac{y}{{\widetilde t} s})}{{\widetilde q} (1 - \frac{1}{{\widetilde q}}) (1 - \frac{y}{s {\widetilde q}})}, \\
\nonumber \\
& \psi_{0,1} = \dfrac{{\widetilde t} (1 - \frac{1}{{\widetilde t}}) (1 - \frac{1}{{\widetilde t} y})}{{\widetilde q} (1 - \frac{1}{{\widetilde q}}) (1 - \frac{1}{y {\widetilde q}})}
\end{align}
\smallskip\\
We actually computed ${\cal B}^{(A)}_{i,j}$ and $\psi_{i,j}$ up to level 4. Based on this information, we propose a relation between the defect partition function and the $q$-deformed conformal block:

\begin{align}
\boxed{ \ \ \ \sum\limits_{i,j = 0}^{\infty} \psi_{i,j} X^i (p/X)^j = \rho(p) \sum\limits_{i,j = 0}^{\infty} {\cal B}^{(A)}_{i,j} x^i (p/x)^j \ \ \ }
\label{5dTorusAGTDefect}
\end{align}
\smallskip\\
under identification of parameters

\begin{align}
\boxed{ \ \ \ \begin{array}{ccc}
X = q^{-\beta N - 1} x, \ \ \ Y = q^{\beta-2+a} \\
\\
{\widetilde t} = q^{-\beta N}, \ \ \ {\widetilde q} = q^{\beta}, \ \ \ s = q^{-1}
\end{array} \ \ \ }
\label{IdentificationDefect}
\end{align}
\smallskip\\
The correction function $\rho(p)$ is independent of $a$ and $x$. Therefore, we verified up to level 4 that defect partition function for circular quiver (Shiraishi function) matches the degenerate torus $q$-conformal block up to $a$- and $x$-independent prefactor. Unlike $\sigma(p,x)$, we do not have a simple conjecture for $\rho(p)$.

\section{Conclusion}

The new dictionary (\ref{5dTorusAGT}) and (\ref{5dTorusAGTDefect}) is interesting in and of itself. This is a non-trivial check of five-dimensional AGT conjecture beyond the simplest case of linear quiver. As a byproduct, we established presence of interesting correction factors $\sigma(p,x)$ and $\rho(p)$. However, potential applications and usefulness of the new formulas go further. We suggest the following short list of questions where the new formulas might be useful:

\paragraph{$\bullet$ Equations satisfied by Shiraishi functions. }It is well-known that degenerate conformal blocks satisfy simple differential equations \cite{BPZ}. In particular, when this conformal block is either 5-point spherical block or 2-point toric block, they can be interpreted as non-stationary differential equations \cite{DefectAGT}.
In the $q$-deformed case, these differential equations are generalized to non-stationary difference equations \cite{qBPZ}. All of these equations are believed to have a simple explanation/origin in the null-vector conditions for degenerate Verma modules of Virasoro or $q$-Virasoro algebra. It would be desirable to provide such clarification for similar equations, satisfied by Shiraishi functions \cite{ShiraishiNonStat}. The integral formula (\ref{5dTorusAGTDefect}) that we suggested is expected to provide a direct connection with null-vector conditions for Shiraishi functions.

\paragraph{$\bullet$ Potential new bases and series expansions. }As explained in \cite{HS}, the integrand of Dotsenko-Fateev integrals can be expanded in various bases of symmetric functions, leading to appealing derivation of AGT conjecture as Hubbard-Stratonovich duality. In the context of (\ref{5dTorusAGT}) and (\ref{5dTorusAGTDefect}), this opens a possibility to expand the integrand in some particularly nice basis and obtain new series expansion for Shiraishi functions. This is interesting to investigate and will be considered elsewhere.

\paragraph{$\bullet$ Elliptic $SL(2,{\mathbb Z})$ identities. }In the ongoing project \cite{EllipticSL2Z}, Shiraishi functions are used to construct matrices $S$ and $T$ that satisfy $SL(2,{\mathbb Z})$ identities. These matrices generalize the modular matrices of refined Chern-Simons theory \cite{rCS} by including two parameters $p,s$ associated with elliptic deformation. It is important to note that the authors of \cite{EllipticSL2Z} relied upon specialization ${\widetilde t}^{2} {\widetilde q}^{K} = 1$ with Chern-Simons level $K = 2$ to prove their results. At the same time, formula (\ref{5dTorusAGTDefect}) is valid for a more general specialization (\ref{IdentificationCircular}), or ${\widetilde t}^{2} {\widetilde q}^{2N} = 1$ with $N = 1,2,3,\ldots$. It is tempting to speculate that integral formula (\ref{5dTorusAGTDefect}) may be useful to construct elliptic $SL(2,{\mathbb Z})$ identities for higher Chern-Simons level $K > 2$.

\paragraph{}After 16 years of development, the AGT relation remains a major topic of interest unifying several areas of research including conformal field theory, supersymmetric gauge theory, topological strings, matrix models, and integrable systems. Higher genus generalizations of AGT are still largely open for investigation. Current paper represents a modest step in this direction for genus 1. It would be very interesting to explore the possibility of $q$-deformation for genus 2 and higher.

\section*{Acknowledgements}

Our work is partly supported by the state assignment of the Institute for Information Transmission Problems
of RAS.


\begin{thebibliography}{27}

\bibitem{AGT} L. Alday, D. Gaiotto and Y. Tachikawa, \emph{Liouville Correlation Functions from Four-dimensional Gauge Theories}, Lett.Math.Phys.91:167-197,2010, https://doi.org/10.1007/s11005-010-0369-5, https://arxiv.org/abs/0906.3219

\bibitem{LMNS1} G.Moore, N.Nekrasov, S.Shatashvili, \emph{Integrating Over Higgs Branches}, Comm.Math.Phys. 209 (2000) 97-121, arxiv.org/abs/hep-th/9712241

\bibitem{LMNS2} A. Losev, N.Nekrasov, S.Shatashvili, \emph{Issues in Topological Gauge Theory}, Nucl.Phys.B534:549-611,1998, arxiv.org/abs/hep-th/9711108

\bibitem{LMNS3} A.Losev, N.Nekrasov, S.Shatashvili, \emph{Testing Seiberg-Witten Solution}, arxiv.org/abs/hep-th/9801061

\bibitem{Nekrasov} N.Nekrasov, \emph{Seiberg-Witten Prepotential From Instanton Counting}, Adv.Theor.Math.Phys.7:831-864,2004, arxiv.org/abs/hep-th/0206161

\bibitem{OPE} A.Marshakov, A.Mironov, A. Morozov, \emph{On Combinatorial Expansions of Conformal Blocks}, Theor.Math.Phys.164:831-852,2010; Teor.Mat.Fiz.164:3-27,2010, arxiv.org/abs/0907.3946

\bibitem{OPEhigher} V.Alba, And.Morozov, \emph{Check of AGT Relation for Conformal Blocks on Sphere}, Nucl.Phys.B840:441-468,2010, arxiv.org/abs/0912.2535

\bibitem{DF} Vl.Dotsenko and V.Fateev, \emph{Conformal algebra and multipoint correlation functions in 2D statistical models}, Nucl.Phys. B240 (1984) 312-348

\bibitem{DFnew} A.Mironov, A.Morozov, Sh.Shakirov, \emph{Conformal blocks as Dotsenko-Fateev Integral Discriminants}, Int.J.Mod.Phys.A25:3173-3207,2010, arxiv.org/abs/1001.0563

\bibitem{DV} R. Dijkgraaf, C. Vafa, \emph{Matrix Models, Topological Strings, and Supersymmetric Gauge Theories}, Nucl.Phys. B644 (2002) 3-20, arxiv.org/abs/hep-th/0206255

\bibitem{Selb} A. Selberg, \emph{Remarks on a multiple integral} (1944), Norsk Mat. Tidsskr. 26: 71–78

\bibitem{Kadell} K. W. J. Kadell, \emph{The Selberg–Jack symmetric functions}, Adv. Math. 130 (1997) 33–102

\bibitem{HS} A.Mironov, A.Morozov, Sh.Shakirov, \emph{A direct proof of AGT conjecture at beta = 1}, JHEP 1102:067,2011, arxiv.org/abs/1012.3137
    
\bibitem{MMNek} A.~Mironov and A.~Morozov,
\emph{Superintegrability as the hidden origin of the Nekrasov calculus},
Phys. Rev. D \textbf{106} (2022) no.12, 126004,
%doi:10.1103/PhysRevD.106.126004
arXiv.org/abs/2207.08242

\bibitem{5dAGT} H. Awata, Y. Yamada, \emph{Five-dimensional AGT Conjecture and the Deformed Virasoro Algebra}, JHEP 1001:125,2010, arxiv.org/abs/0910.4431

\bibitem{5dAGT2} A.Mironov, A.Morozov, Sh.Shakirov, A.Smirnov, \emph{Proving AGT conjecture as HS duality: extension to five dimensions}, Nucl.Phys. B855 (2012) 128-151, arxiv.org/abs/1105.0948

\bibitem{DELL} P. Koroteev, Sh. Shakirov, \emph{The Quantum DELL System}, Lett. Math. Phys. 110 (2020)969-999, arxiv.org/abs/1906.10354

\bibitem{MMZ} A.~Mironov, A.~Morozov and Y.~Zenkevich,
\emph{Duality in elliptic Ruijsenaars system and elliptic symmetric functions},
Eur. Phys. J. C \textbf{81} (2021) no.5, 461,
%doi:10.1140/epjc/s10052-021-09248-9
arxiv.org/abs/2103.02508

\bibitem{MMdell} A.~Mironov and A.~Morozov,
\emph{On the status of DELL systems},
Nucl. Phys. B \textbf{999} (2024), 116448,
%doi:10.1016/j.nuclphysb.2024.116448
arxiv.org/abs/2309.06403

\bibitem{WnAGT} N. Wyllard, \emph{$A_{N-1}$ conformal Toda field theory correlation functions from conformal ${\cal N}=2 SU(N)$ quiver gauge theories}, JHEP 0911:002,2009, arxiv.org/abs/0907.2189
    
\bibitem{MMWn} A.~Mironov and A.~Morozov,
\emph{On AGT relation in the case of U(3)},
Nucl. Phys. B \textbf{825} (2010), 1-37,
%doi:10.1016/j.nuclphysb.2009.09.011
arxiv.org/abs/0908.2569

\bibitem{DefectAGT} A.Marshakov, A.Mironov, A.Morozov, \emph{On AGT Relations with Surface Operator Insertion and Stationary Limit of Beta-Ensembles}, J.Geom.Phys.61:1203-1222,2011, arxiv.org/abs/1011.4491

\bibitem{BPZ} A. Belavin, A. Polyakov, and A.B. Zamolodchikov, \emph{Infinite conformal symmetry in two-dimensional quantum field theory} (1984) Nuclear Physics B. 241 (2): 333–380

\bibitem{ShiraishiNonStat} J. Shiraishi, \emph{Affine Screening Operators, Affine Laumon Spaces, and Conjectures Concerning Non-Stationary Ruijsenaars Functions}, arxiv.org/abs/1903.07495

\bibitem{qBPZ} Sh.Shakirov, \emph{Non-stationary difference equation for q-Virasoro conformal blocks}, Lett.Math.Phys. 114 (2024) 115,  arxiv.org/abs/2111.07939

\bibitem{qPain} H.Awata, K.Hasegawa, H.Kanno, R.Ohkawa, Sh.Shakirov, J.Shiraishi, Y.Yamada, \emph{Non-Stationary Difference Equation and Affine Laumon Space: Quantization of Discrete Painlevé Equation}, SIGMA 19 (2023), 089, arxiv.org/abs/2211.16772

\bibitem{qBPZ2} H.Awata, K.Hasegawa, H.Kanno, R.Ohkawa, Sh.Shakirov, J.Shiraishi, Y.Yamada, \emph{Non-Stationary Difference Equation and Affine Laumon Space II: Quantum Knizhnik-Zamolodchikov Equation}, SIGMA 20 (2024), 077, arxiv.org/abs/2309.15364

\bibitem{qBPZ3} H.Awata, K.Hasegawa, H.Kanno, R.Ohkawa, Sh.Shakirov, J.Shiraishi, Y.Yamada, \emph{Non-stationary difference equation and affine Laumon space III : Generalization to $\widehat{\mathfrak{gl}}_N$}, arxiv.org/abs/2510.27142
    
\bibitem{ELS} H.~Awata, H.~Kanno, A.~Mironov and A.~Morozov,
\emph{Elliptic lift of the Shiraishi function as a non-stationary double-elliptic function},
JHEP \textbf{08} (2020), 150,
%doi:10.1007/JHEP08(2020)150
arXiv.org/abs/2005.10563

\bibitem{diamond} A.~Mironov, A.~Morozov, A.~Popolitov and Z.~Zakirova,
\emph{Diamond of triads},
Pisma Zh. Eksp. Teor. Fiz. \textbf{121} (2025) no.9, 788-795,
%doi:10.31857/S0370274X25050127
arXiv.org/abs/2503.07592

\bibitem{Toric} A.Mironov, A.Morozov, Sh.Shakirov, \emph{On "Dotsenko-Fateev" representation of the toric conformal blocks}, J.Phys.A44:085401,2011, arxiv.org/abs/1010.1734

\bibitem{AGTToric} V.Alba, And.Morozov, \emph{Non-conformal limit of AGT relation from the 1-point torus conformal block}, JETP Lett.90:708-712,2009, arxiv.org/abs/0911.0363

\bibitem{ToricGeneral} K.Maruyoshi, F.Yagi, \emph{Seiberg-Witten curve via generalized matrix model}, JHEP 1101:042,2011, arxiv.org/abs/1009.5553

\bibitem{HigherGenusGeneral} G.Bonelli, K.Maruyoshi, A.Tanzini, F.Yagi, \emph{Generalized matrix models and AGT correspondence at all genera}, JHEP 1107:055,2011, arxiv.org/abs/1011.5417

\bibitem{q-Virasoro}  J.Shiraishi, H.Kubo, H.Awata, S.Odake, \emph{A Quantum Deformation of the Virasoro Algebra and
the Macdonald Symmetric Functions}, Lett.Math.Phys. 38 (1996) 33, doi.org/10.1007/BF00398297,
arxiv.org/abs/q-alg/9507034

\bibitem{EllipticSL2Z} S.Arthamonov, Sh.Shakirov, \emph{An Elliptic Generalization of $A_1$ Spherical DAHA at $K=2$}, IMRN 19(2024)13046–13084, arxiv.org/abs/2306.00215

\bibitem{rCS} M.Aganagic, Sh.Shakirov, \emph{Knot Homology from Refined Chern-Simons Theory}, Comm.Math.Phys. 333 (2015) 1, 187-228, arxiv.org/abs/1105.5117

\end{thebibliography}
\end{document}